\documentstyle[12pt]{article}

\newlength{\extraspace}
\newlength{\extraspaces}
\catcode`\@=11
 
\def\numberbysection{\@addtoreset{equation}{section}
\def\theequation{\arabic{section}.\arabic{equation}}}

\begin{document}
%
\thispagestyle{empty}
\begin{center}
\begin{flushright}
TIT/HEP--386 \\
{\tt hep-th/9802004} \\
February, 1998 \\
\end{flushright}
\vspace{3mm}
\begin{center}
{\Large
{\bf M theory and the ``integrating in'' method \\
 with an antisymmetric tensor
}} 
\\[18mm]

{\sc Hodaka~Oda},\footnote{
\tt e-mail: hoda@th.phys.titech.ac.jp} \hspace{2.0mm}
{\sc Shigemitsu~Tomizawa},\footnote{
\tt e-mail: stomizaw@th.phys.titech.ac.jp} \hspace{2.0mm}
{\sc Norisuke~Sakai}\footnote{
\tt e-mail: nsakai@th.phys.titech.ac.jp} \hspace{2.0mm}
and \hspace{2.0mm}
{\sc Tadakatsu~Sakai}\footnote{
\tt e-mail: tsakai@th.phys.titech.ac.jp} \\[3mm]
{\it Department of Physics, Tokyo Institute of Technology \\[2mm]
Oh-okayama, Meguro, Tokyo 152-8551, Japan} \\[4mm]

\end{center}
\vspace{18mm}
{\bf Abstract}\\[5mm]
{\parbox{13cm}{\hspace{5mm}
%
Recently, a non-hyperelliptic curve describing 
the Coulomb branch of $N=2$ SUSY $SU(N_c)$ Yang-Mills theory with
an antisymmetric tensor matter was proposed using a configuration of a
single 
M theory five-brane. We study the singular surface in the moduli space
of the curve 
to compare it with results from the ``integrating in'' method in field
theory. 
In order to achieve the consistency, we find it necessary to take
account of 
an additional superpotential $W_{\Delta}$ which has been neglected so
far. 
The explicit form of $W_{\Delta}$ is worked out. 
}}
\end{center}
\vfill
\newpage
\vfill
\newpage
\addtocounter{section}{1}
\setcounter{equation}{0}
\setcounter{section}{0}
\setcounter{footnote}{0}
\numberbysection

\vspace{7mm}
\pagebreak[3]
\addtocounter{section}{1}
\setcounter{equation}{0}
\setcounter{subsection}{0}
\setcounter{footnote}{0}
\begin{center}
{\large {\bf \thesection. Introduction }}
\end{center}
\nopagebreak
\medskip
\nopagebreak
\hspace{3mm}

In recent years, deeper understanding of supersymmetric (SUSY) gauge 
theories in various dimensions has been
gained by realizing them on the world-volumes of 
D-branes \cite{hw}-\cite{EGKT}.
Witten \cite{witten} also pointed out that 
intersecting brane configurations of Type IIA string theory
corresponding to 
$N=2$ SUSY gauge theories in four dimensions can be described by a
single
M theory five-brane wrapping around a Riemann surface.
The Riemann surface is nothing but the Seiberg-Witten curve \cite{sw} 
and therefore the five-brane configuration contains the structure of the 
moduli space of vacua. The M theoretic method is also applied to discuss 
various aspects of SUSY gauge theories \cite{15}-\cite{deHOO2} 
and found to be quite useful to understand them.

On the other hand, field-theoretic approaches also provide us with 
important informations on the Seiberg-Witten curves. 
One of them is based on the deformation to $N=1$ SUSY. 
The moduli space of the $N=2$ SUSY vacua in the Coulomb phase 
exhibits singularities where solitons 
such as monopoles or dyons become massless. 
When $N=2$ SUSY gauge theories are broken to $N=1$ SUSY by perturbations
of 
tree-level superpotentials, 
only these singularities remain as $N=1$ SUSY vacua \cite{sw}. 
Conversely, we can tune parameters of superpotentials in 
$N=1$ SUSY Yang-Mills theories with an adjoint matter 
field in order to obtain $N=2$ SUSY Yang-Mills theories. 
By this procedure, one expects that the singularity surfaces in the
$N=2$ moduli space 
can be reached. 
Thus, by studying the low energy effective action of $N=1$ Yang-Mills
theory 
with an adjoint matter field with a tree-level superpotential chosen
properly, we can derive some informations on the singular surface of 
the $N=2$ moduli spaces. 
In fact, Elitzur et al.~have developed a method to obtain the
singularity 
surfaces in the $N=2$ SUSY Yang-Mills theories by using a single
confined 
photon in the $N=1$ SUSY gauge theories \cite{EFGIR}. 
In this way, the curve of the $N=2$ SUSY theory can be recovered 
by ``integrating in'' \cite{Intriligator} the adjoint matter 
fields in the $N=1$ low energy effective theory. 
This ``integrating in'' method 
has been extended to SUSY Yang-Mills theories with various gauge groups 
including exceptional groups \cite{TY1}-\cite{TY2}. 
In general, however, the effective superpotential is not completely
fixed by symmetries and holomorphy.
Possible additional terms are usually denoted as $W_\Delta$. 
In these ``integrating in'' approaches, a crucial assumption has been
made: 
the low energy effective superpotential has a minimal form, namely 
 $W_\Delta = 0$. So far, this has provided us with 
consistent results.

Recently, Landsteiner and Lopez \cite{LL1} have proposed a
non-hyperelliptic 
curve describing the Coulomb branch of $N=2$ SUSY $SU(N_c)$ Yang-Mills
theory with
an antisymmetric tensor matter from a configuration of a single 
M theory five-brane. 
Although the proposed curve passes some consistency checks, it seems 
necessary to make sure of it further from other points of view. 
The purpose of this paper is to obtain the singularity surface of the 
$N=2$ SUSY  $SU(N_c)$ Yang-Mills theory with 
an antisymmetric tensor matter by using the ``integrating in'' method. 
We find that the usual ``integrating in'' method assuming $W_{\Delta}=0$ 
gives a singularity surface which disagrees with the proposed M
theoretic curve. 
By assuming that the brane configuration is correct, 
we find that there exists a nontrivial $W_{\Delta} \not=0$ which gives a
singular 
surface consistent with the M theoretic curve at least up to certain
high powers of the 
dynamical scale $\Lambda$ of the gauge interactions. 
Our results can be regarded as an evidence for the necessity of
nontrivial 
$W_{\Delta}$.

Section 2 gives a brief review of the brane configuration in M theory 
describing the $N=2$ SUSY $SU(N_c)$ Yang-Mills theory with
an antisymmetric tensor matter. 
In section 3, we discuss the singular surface of the moduli space
assuming 
$W_\Delta = 0$.
It is shown that the singular surface is inconsistent with 
the M theoretic curve obtained from the brane configuration.
In section 4, we 
derive the explicit form of $W_\Delta \not=0$ by requiring the
consistency 
of the singular surface with the M theoretic curve. 
Section 5 contains a discussion.

\vspace{7mm}
\pagebreak[3]
\addtocounter{section}{1}
\setcounter{equation}{0}
\setcounter{subsection}{0}
\addtocounter{footnote}{0}
\begin{center}
{\large {\bf \thesection. The brane configuration}}
\end{center}
\nopagebreak
\medskip
\nopagebreak
\hspace{3mm}

In this section we briefly review the brane configuration in M theory 
describing the Coulomb branch of the $N=2$ SUSY $SU(N_c)$ Yang-Mills
theory with 
an antisymmetric tensor matter \cite{LL1}. 
Let us first examine the brane configuration in the type IIA string
picture. 
Consider type IIA string theory in flat space-time where
$x^0$ denotes the time coordinate and
$x^1,\ldots,x^9$ denote the space coordinates.
The brane configuration consists of
an orientifold sixplane of charge $-4$
with the world-volume coordinates
($x^0, x^1, x^2, x^3, x^7, x^8, x^9$),
Neveu Schwarz (NS) 5 branes with the world-volume coordinates
($x^0, x^1, x^2, x^3, x^4, x^5$),
and Dirichlet (D) 4 branes with the world-volume coordinates
($x^0, x^1, x^2, x^3, x^6$).
The orientifold sixplane sits at
$x^4=x^5=x^6=0$.
This means that the space-time should be identified under the 
transformation
\begin{eqnarray}
(x^4, x^5, x^6) \to (-x^4, -x^5, -x^6).
\end{eqnarray}
One NS5 brane is placed on top of the orientifold sixplane
and
the other NS5 brane is to the right of it.
Further there are $N_c$ D4 branes stretching in between the NS5 branes.
In the left of the orientifold sixplane
we have of course the mirror image of
these branes.
The D4 branes have a finite extent in the $x^6$ direction. 
The four dimensional $N=2$ SUSY gauge
theory we discuss is defined 
on the world-volume coordinates ($x^0, x^1, x^2, x^3$) of the D4 branes.
When all $N_c$ D4 branes coincide, the open strings connecting $N_c$ D4
branes 
in the left of the orientifold six-plane give the $SU(N_c)$ gauge vector
multiplets. 
The open strings connecting the left and right D4 branes give 
a hypermultiplet in the antisymmetric representation of the gauge group 
because of the presence of the orientifold sixplane.

The brane configuration can be reinterpreted in M theory
as a configuration of a single five-brane 
embedded in the eleven-dimensional space-time
${\bf R}^7 \times S$ where $S$ is the Atiyah-Hitchin space 
\cite{AH}, \cite{SW2}.
The ${\bf R}^7$ spans the 0123789 directions,
while $S$ spans the 456 directions in the Type IIA limit 
and wraps around the circle in the eleventh direction $x^{10}$ whose 
radius is denoted by $R$. 
The five-brane world-volume becomes ${\bf R}^4 \times \Sigma$
where ${\bf R}^4$ spans the 0123 directions 
while $\Sigma$ is a curve embedded in the Atiyah-Hitchin space $S$ 
whose complex structure is represented as 
$x y = \Lambda^{2N_c+4} v^{-4}$,
where $v=x^4+i x^5$.
For large $y$ with $x$ fixed, $y$ tends to
$t=\exp (-(x^6+i x^{10})/R )$,
while for large $x$ with $y$ fixed we have
$x \sim t^{-1}$.
In the M theoretic brane configuration, $\Lambda$ represents 
the mass scale corresponding to the dynamical scale of gauge interaction 
in field theory. 
The curve $\Sigma$ is not hyper-elliptic, contrary to the case of the
$N=2$ SUSY QCD 
where matter hypermultiplets are only in the fundamental
representations.

Since there are three NS5 branes involved, the M theoretic curve
describing the 
brane configuration becomes cubic in $y$. 
By using symmetry under $x \leftrightarrow y$, $v \leftrightarrow -v$
and 
other arguments, 
Landsteiner and Lopez has found the following curve $\Sigma$ for 
the above brane configuration in M theory, and proposed it to describe 
the $N=2$ SUSY $SU(N_c)$ Yang-Mills gauge theory with an antisymmetric
tensor 
matter field \cite{LL1} 
\begin{eqnarray}
y^3 + y^2 \left( p(v)+3 \Lambda^{N_c+2} v^{-2}  \right)
+ y \Lambda^{N_c+2} v^{-2} \left( q(v) +3 \Lambda^{N_c+2} v^{-2} \right)
+ \Lambda^{3N_c+6} v^{-6} = 0,
\label{curveold}
\end{eqnarray}
where
\begin{eqnarray}
p(v) = \prod^{N_c}_{i=1}(v-a_i), \qquad {\rm and} \qquad
q(v) = p(-v).
\end{eqnarray}
The $N_c$ parameters $a_i$ represent the positions
of the D4 branes in the IIA string picture.

We denote by $\Phi$ an $N=1$ chiral superfield in the adjoint
representation 
in the $SU(N_c)$ gauge group. 
Together with the $N=1$ vector multiplet $V$ in the adjoint 
representation, it forms an $N=2$ vector multiplet. 
In addition to them, 
we have an antisymmetric tensor matter $A^{ij}$ and
its conjugate $\tilde{A}_{ij}$, 
where $i, j = 1, 2 \ldots ,N_c$ are color indices. 
Both of them are the $N=1$ chiral superfields and 
form together an $N=2$ hypermultiplet.
The tree level 
superpotential $W_{\rm tree}$ contains a 
tree level mass parameter $m$ of the antisymmetric tensor 
matter 
\begin{equation}
W_{\rm tree} = \sqrt{2}
\sum_{i < j \atop k<l }
\tilde{A}_{ij} 
\left( \Phi^i{}_k \delta^j{}_{l} + \delta^i{}_k \Phi^j{}_l 
+ m \delta^i{}_k  \delta^j{}_{l} \right) A^{kl} .
\label{treesp}
\end{equation}

The distance between the average position of the 
D4 branes on the left and the average position of the D4 branes
on the right is equal to the tree level mass parameter $m$ 
of the antisymmetric tensor matter
\begin{eqnarray}
m = \frac{2}{N_c} \sum_{i=1}^{N_c} a_i.
\end{eqnarray}
The distance between the position of each D4 brane and
the average position of the D4 branes on the left corresponds to
 the vacuum expectation value (VEV) of the diagonal element $\phi_i$ of 
the adjoint matter $\Phi$
\begin{eqnarray}
a_i = \frac{m}{2} + 
\langle \phi_i  \rangle,\qquad (i=1,\ldots,N_c),
\end{eqnarray}
where VEV is denoted by $\langle \  \rangle$. 

Rescaling and shifting $y \to ( y - \Lambda^{N_c+2} v^{-1}) v^{-1}$,
the curve (\ref{curveold}) becomes $f(y,v) =0$ where 
\begin{eqnarray}
f(y,v) \equiv
y^3+y^2 v \, p(v) +
y \Lambda^{N_c+2} (q(v) -2 p(v))+
\Lambda^{2N_c+4} v^{-1} \left( p(v)-q(v) \right).
\label{curve}
\end{eqnarray}
Notice that $q(v)=p(-v)$ and 
then $ v^{-1} \left( p(v)-q(v) \right) $ 
has no negative powers in $v$.


Although we do not know any field-theoretical method to obtain the curve
for 
the case involving the antisymmetric tensor matter field, 
we can obtain rich informations on the singular surface of the curve 
describing the Coulomb branch of the  $N=2$ SUSY Yang-Mills gauge
theories 
by the method of ``integrating in'' \cite{EFGIR}, \cite{Intriligator}. 
We will here compute the singular surface of 
the proposed curve, where the discriminant vanishes. 
Since the antisymmetric tensor representation in the $SU(3)$ gauge group
is nothing 
but the antifundamental representation, we shall take the $SU(4)$ gauge
group as 
the simplest nontrivial case. 
Using Maple, in order to obtain the 
discriminant
of the curve for the $SU(4)$ Yang-Mills theory with an 
antisymmetric matter,
we calculate
\begin{eqnarray}
\prod_{i=1} f(y_i,v_i),
\end{eqnarray}
where $(y_i,v_i)$ are solutions of simultaneous equations
\begin{eqnarray}
\frac{\partial f}{\partial y} (y,v)
&\!\!\! = &\!\!\! 0, \\
\frac{\partial f}{\partial v} (y,v)
&\!\!\! = &\!\!\! 0.
\end{eqnarray}
To perform an explicit calculation, we take the $m=0$ case. 
The discriminant\footnote{
Note that the antisymmetric representation of $SU(4)$ is
equivalent to the defining representation of $SO(6)$, for which 
the Seiberg-Witten curve has been derived.
It turns out that the discriminant of the Seiberg-Witten curve for the 
$SO(6)$ with the defining representation contains the factor 
$\Delta_{\rm massive}$ which reduces to $s_3^2 \Delta$ in (\ref{Delta1}) for the massless case. 
}
is found to be
\begin{eqnarray}
{s_3}^{4} {\Delta}^2 \Delta_{\rm unphys},
\label{disc}
\end{eqnarray}where
\begin{eqnarray}
&\!\!\!
\Delta
= &\!\!\!
191102976 {s_3}^{2}\Lambda^{18}
+(-1327104 s_4 {s_2}^{4}
+5308416 {s_2}^{2}{s_4}^{2}
+110592 {s_2}^{6}
-8957952 {s_3}^{4}
\nonumber \\ &\!\!\! &\!\!\!{}
-39813120 s_2 s_4 {s_3}^{2}
-7077888 {s_4}^{3}
+9068544 {s_3}^{2}{s_2}^{3}){\Lambda}^{12}
+(417792 {s_4}^{2}{s_2}^{5}
-1146880 {s_4}^{3}{s_2}^{3}
\nonumber \\ &\!\!\! &\!\!\!{}
+139968 {s_3}^{6}
+245376 {s_3}^{4}{s_2}^{3}
+4096 {s_2}^{9}
-488448 {s_2}^{4}s_4 {s_3}^{2}
+59904 {s_3}^{2}{s_2}^{6}
\nonumber \\ &\!\!\! &\!\!\!{}
+2211840 {s_4}^{3}{s_3}^{2}
+442368 {s_4}^{2}{s_2}^{2}{s_3}^{2}
-67584 s_4 {s_2}^{7}
+1179648 {s_4}^{4}s_2
+124416 s_2 s_4 {s_3}^{4})\Lambda^6
\nonumber \\ &\!\!\! &\!\!\!{}
-27648 {s_4}^{2}{s_3}^{4}{s_2}^{2}
-5632 {s_4}^{2}{s_3}^{2}{s_2}^{5}
-73728 s_2 {s_4}^{4}{s_3}^{2}
+128 {s_3}^{2}s_4 {s_2}^{7}
-256 {s_4}^{2}{s_2}^{8}
\nonumber \\ &\!\!\! &\!\!\!{}
-16 {s_3}^{4}{s_2}^{6}
+65536 {s_2}^{2}{s_4}^{5}
+38912 {s_4}^{3}{s_2}^{3}{s_3}^{2}
+7776 {s_3}^{6}s_4 s_2
-24576 {s_2}^{4}{s_4}^{4}
-729 {s_3}^{8}
\nonumber \\ &\!\!\! &\!\!\!{}
+2016 {s_3}^{4}s_4 {s_2}^{4}
+13824 {s_4}^{3}{s_3}^{4}
-65536 {s_4}^{6}
-216 {s_3}^{6}{s_2}^{3}
+4096 {s_4}^{3}{s_2}^{6} ,
\label{Delta1} \\
&\!\!\!
\Delta_{\rm unphys}
\!\!\!\!\!\!
&\!\!\!
\,\,\,\,\,
=
(-{s_4}^3+27 \Lambda^{6} {s_3}^2).
\end{eqnarray}
Here $s_i$ are the moduli parameters,
\begin{eqnarray}
\langle {\rm det} (x-\Phi) \rangle = x^{N_c} + \sum^{N_c}_{i=2}
x^{N_c-i} s_i.
\label{moduli}
\end{eqnarray}
The factor $\Delta_{\rm unphys}$ is believed to be 
unphysical \cite{LL1}. 
On the other hand, the factor ${s_3}^4$
exhibits a singularity expected for
the 
massless antisymmetric tensor matter field in the classical limit 
$(\Lambda \to 0)$, 
as can be seen from the
tree-level 
superpotential (\ref{treesp}) 
\begin{equation}
{s_3}^2 = \langle \prod_{i>j}(\phi_i + \phi_j) \rangle .
\end{equation}
It is interesting to observe that this singularity is identical to the
classical limit 
$(\Lambda \rightarrow 0)$ even though we are not restricted to the weak
coupling 
case. 
In order to see the singularity associated with the massless gauge
fields, 
we shall take the classical limit ($\Lambda \to 0$). 
Then the factor $\Delta$ becomes 
\begin{eqnarray}
\Delta
&\!\!\! \to &\!\!\!
 \langle \prod_{i>j}(\phi_i - \phi_j)^4 \rangle .
\end{eqnarray}
This is nothing but the classical singularity 
where the non-Abelian gauge symmetry is enhanced.
We conclude that ${s_3}^2 \Delta$ correctly reproduces the singularities 
in the classical limit. 
%

\vspace{7mm}
\pagebreak[3]
\addtocounter{section}{1}
\setcounter{equation}{0}
\setcounter{subsection}{0}
\addtocounter{footnote}{0}
\begin{center}
{\large {\bf \thesection. The ``integrating in'' method
}}
\end{center}
\nopagebreak
\medskip
\nopagebreak
\hspace{3mm}

In this section, we analyze the singular surface in the moduli space of
the 
Coulomb branch by using the ``integrating in'' method 
in the field-theoretic framework. 
This method enables us to gain informations on 
the singular surface in the 
Coulomb branch 
taking into account of nonperturbative quantum effects.

The $N=2$ SUSY is broken to $N=1$ 
by adding a perturbation $\Delta W$
to the tree-level $N=2$ superpotential $W_{\rm tree}$ in (\ref{treesp}) 
\begin{eqnarray}
\Delta W = \sum_{k=2}^{N_c} \frac{g_k}{k} {\rm Tr} (\Phi^k).
\label{3.1}
\end{eqnarray}
The classical VEV's of $\Phi$ 
are obtained from the classical equations of motion, 
$\partial ( W_{\rm tree}+\Delta W ) / \partial \Phi=0$, 
and similarly for $A^{ij},  \tilde{A}_{ij}$. 
We are interested in the Coulomb branch, where 
${A}^{ij}=\tilde{A}_{ij}=0$. 
After $SU(N_c)$ rotations, the generic VEV can be reduced to 
$\Phi_{\rm cl} = {\rm diag}(M,M,M_3,M_4, \ldots, M_{N_c})$, 
where $M=g_{N_c-1}/g_{N_c}$.
In that case, the gauge group $SU(N_c)$ is broken to 
$SU(2) \times U(1)^{N_c-2}$. 
The nonperturbative effects due to the gaugino condensation 
of the $SU(2)$ super Yang-Mills theory provides the additional 
superpotential 
\begin{eqnarray}
W_d = \pm 2 g_{N_c} \left(
\Lambda_{\rm FT}^{N_c+2} G \right)^{1/2}.
\label{gauginocond}
\end{eqnarray}
where 
\begin{eqnarray}
G = \prod_{p=3}^{N_c}(M_p+M+m),
\end{eqnarray}
and
$\Lambda_{\rm FT}$ is the dynamical scale of the $SU(N_c)$ gauge theory
with an antisymmetric matter in field theory.
The $\Lambda_{\rm FT}$ must be proportional to $\Lambda$ of the M theory 
brane configuration in the previous section:
\begin{eqnarray}
\Lambda_{\rm FT} = c \Lambda \qquad c \in {\bf C}, 
\label{renscc}
\end{eqnarray}
where $c$ is a renormalization-scheme-dependent 
constant. 
Following Elitzur et.~al.~\cite{EFGIR}, 
we obtain the low-energy effective superpotential 
for the $N=1$ super Yang-Mills theory 
\begin{eqnarray}
W_L = \Delta W(\Phi=\Phi_{\rm cl}(g_k))+
W_d
+ W_\Delta, 
\end{eqnarray}
where $W_{\Delta}$ is a possible additional superpotential 
constrained only by holomorphy and symmetry \cite{Intriligator}. 

The VEV of gauge invariants can be defined as 
\begin{eqnarray}
\langle u_k \rangle =\langle  \frac{1}{k} {\rm Tr}(\Phi^k) \rangle,
\end{eqnarray}
which 
are obtained by differentiating $W_L$,
\begin{eqnarray}
\langle u_k \rangle = \frac{\partial W_L}{\partial g_k}.
\label{VEV}
\end{eqnarray}
The Seiberg-Witten curve must be singular
when $u_k=\langle u_k \rangle$.
The VEV's are related to the moduli parameters $s_i$ in
eq.(\ref{moduli}) 
through the Newton formula 
\begin{eqnarray}
k s_k = - \sum_{j=1}^{k}j s_{k-j}\langle u_j \rangle,
\end{eqnarray}
with $s_0=1$, $s_1=0$.

As a simplest explicit example, we consider the $SU(4)$ case. 
Then, the VEV's and the gaugino condensation 
are given in terms of coupling parameters in 
the superpotential as 
\begin{eqnarray}
M
\!\!\!& =\!\!\!& z_3 \\
M_3
\!\!\!& =\!\!\!& -z_3+\sqrt{-{z_3}^2-z_2} \\
M_4
\!\!\!& =\!\!\!& -z_3-\sqrt{-{z_3}^2-z_2} \\
G
\!\!\!& =\!\!\!& m^2+{z_3}^2+z_2,
\end{eqnarray}
where $z_3$ and $z_2$ are complex parameters
\begin{eqnarray}
z_3 \equiv \frac{g_3}{g_4}, \qquad
z_2 \equiv \frac{g_2}{g_4}.
\end{eqnarray}

In the remainder of the paper, 
we will 
explore the singular surface by using the ``integrating in'' 
method, in order to compare it with curve obtained from the 
M theory five-brane. 
First, we assume $W_\Delta=0$ in this section. 
Then we find the VEV including quantum effects using 
eq.(\ref{VEV}) as 
\begin{eqnarray}
\langle u_2 \rangle
\!\!\!& =\!\!\!& 
{z_3}^2-z_2 \pm \Lambda_{\rm FT}^3 G^{-1/2}
\nonumber \\
\langle u_3 \rangle
\!\!\!& =\!\!\!& 
2 {z_3}^3 + 2 z_3 z_2 \pm 2 z_3 \Lambda_{\rm FT}^3 G^{-1/2}
\nonumber \\
\langle u_4 \rangle
\!\!\!& =\!\!\!& 
-\frac{3}{2} {z_3}^4 -2 {z_3}^2 z_2 +\frac{1}{2} {z_2}^2
\pm (2 m^2 + z_2) \Lambda_{\rm FT}^3 G^{-1/2}.
\label{oldsingular}
\end{eqnarray}
These relations define a codimension-one surface in the moduli space. 
It should correspond to the singular surface 
of the proposed curve (\ref{curve}) for $SU(4)$ with an antisymmetric 
tensor matter, namely the vanishing discriminant of the curve. 
We will find, however, the discriminant of the curve (\ref{curve}) does not 
vanish on $u_k=\langle u_k \rangle$ in eq.(\ref{oldsingular}) for the case 
$m=0$.

We shall now test if the discriminant vanishes 
for any values of coupling parameters $z_3$ and $z_2$ 
by choosing an appropriate value for the renormalization-scheme-dependent 
factor $c$ in eq.(\ref{renscc}). 
We find that the factor $\Delta$ in the discriminant (\ref{disc}), 
for instance, 
becomes on the codimension-one surface (\ref{oldsingular}) for $m=0$ 
\begin{eqnarray}
\lefteqn{\Delta(u_k = \langle u_k \rangle)} \nonumber \\
&\!\!\! = &\!\!\! -1024(c^6-4)
\left({z_3}^2+{z_2}\right)^3
\left( 5 {z_3}^2+{z_2} \right)^6
\Lambda^6 \nonumber \\
&\!\!\! &\!\!\! {}
+
\left(
(811 c^6 -3744) {z_3}^6 +(117 c^6-768){z_3}^4{z_2}
+(-327 c^6+ 1248){z_3}^2{z_2}^2+(47c^6-192){z_2}^3
\right)
\nonumber \\ &\!\!\! &\!\!\! {} \times
512
\left({z_3}^2+{z_2}\right)^{3/2}
\left( 5 {z_3}^2+{z_2} \right)^3
c^3 \Lambda^{9}
+ {\cal O}(\Lambda^{12}) 
\label{skintoDelta1} \label{3.15} \\
&\!\!\! \not\equiv &\!\!\! 0 \nonumber 
\end{eqnarray}
To be more precise,
there is no complex number $c$ satisfying
$\Delta(u_k = \langle u_k \rangle) \equiv 0$
for any values of $z_3$ and $z_2$. 
Therefore the discriminant of the curve (\ref{curve}) does not vanish 
on the codimension-one surface (\ref{oldsingular}) obtained by 
assuming $W_{\Delta}=0$ in the ``integrating in'' method. 
Although we have no rigorous means to test the curve (\ref{curveold}) 
obtained in M theory for general $N_c$, we are confident 
that the curve is correct at least for $N_c=4$, 
since we have checked 
that the discriminant agrees with that of $SO(6)$ with the defining 
representation.
Therefore we conclude that
the assumption $W_\Delta=0$ in the case of the 
$SU(4)$ theory with an antisymmetric tensor matter leads us to 
inconsistent results and that the assumption $W_\Delta=0$ is 
not correct. 

\vspace{7mm}
\pagebreak[3]
\addtocounter{section}{1}
\setcounter{equation}{0}
\setcounter{subsection}{0}
\addtocounter{footnote}{0}
\begin{center}
{\large {\bf \thesection. Non-zero $W_\Delta$
}}
\end{center}
\nopagebreak
\medskip
\nopagebreak
\hspace{3mm}

In the previous section, we found 
that the M theory curve (\ref{curve}) is inconsistent with 
the codimension-one surface obtained as a candidate for 
the singular surface in the moduli space assuming $W_\Delta=0$ 
in the ``integrating in'' method. 
In this section, we discuss the possibility of non-zero $W_\Delta$ 
instead.

We first note that, in the classical limit ($\Lambda \rightarrow 0$), 
the discriminant of the M theory curve (\ref{curve}) vanishes on 
the codimension-one surface (\ref{oldsingular}) obtained by assuming 
$W_{\Delta}=0$ in the ``integrating in'' method. 
Eq.(\ref{skintoDelta1}) shows that the discriminant of the curve 
vanishes on the co-dimension-one surface in the leading order 
of $\Lambda$, i.e. 
up to order $\Lambda^6$, 
provided $c^6=4$, 
\begin{eqnarray}
\Lambda^6_{\rm FT}=4 \Lambda^{6}, 
\qquad 
W_d =\pm 4 g_{4} \Lambda^3 G^{1/2} .
\end{eqnarray}
Now we wish to explore to higher orders of $\Lambda$ 
whether we can find a nontrivial $W_{\Delta}$ 
which provides the singular surface consistent with the 
vanishing discriminant of the curve.
Since the right-hand-side of eq.(\ref{3.15}) 
consists of terms with integer powers of $\Lambda^3$, we need to 
introduce terms with $\Lambda^{3n}$ $n \in {\bf N}$ only
\begin{eqnarray}
W_\Delta = 
\sum_{k=2}^{\infty} C_k (\Lambda^3)^k.
\label{WDeltaCi}
\end{eqnarray}
Since the $\Lambda^3$ term is given by $W_d$, 
we assume $k \geq 2$.

The additional superpotential $W_\Delta$ must satisfy the 
following conditions \cite{Intriligator} 
\begin{eqnarray}
W_\Delta \to 0 & & \mbox{as} \hspace{0.3cm} g_2 \to \infty \quad \mbox{with} 
\quad g_4 \Lambda^3 G^{1/2} \quad \mbox{fixed}, 
\nonumber \\
W_\Delta \to 0 & & \mbox{as} \hspace{0.3cm} \Lambda \to 0,
\label{WDeltacondition}
\end{eqnarray}
and carry charge $(2,2)$ under $U(1)_R \times U(1)_J$.

We list below the charge and mass dimension of the parameters.
\begin{eqnarray}
\begin{tabular}{cccc} 
         &$U(1)_R$&$U(1)_J$ & {\rm Dimension}\\
$g_2$    &   $-2$ &   $2$   & $1$  \\
$g_3$    &   $-4$ &   $2$   & $0$  \\
$g_4$    &   $-6$ &   $2$   & $-1$ \\
$z_3$    &   $ 2$ &   $0$   & $1$  \\
$z_2$    &   $ 4$ &   $0$   & $2$  \\
$\Lambda$&   $ 2$ &   $0$   & $1$  \\
$G$      &   $ 4$ &   $0$   & $2$  \\
$m$      &   $ 2$ &   $0$   & $1$ 
\end{tabular}
\end{eqnarray}
{}From this table, we see 
that $W_\Delta$ can be represented as
\begin{eqnarray}
W_\Delta = g_4 \sum_{k=2}^{\infty}
f_{k}(z_3, z_2) \Lambda^{3k},
\end{eqnarray}
where $f_{k}(z_3, z_2)$ is any function 
which carries the charge ($4-6k,0$).
Note that we are discussing $SU(4)$ Yang-Mills theory with a
massless ($m=0$) antisymmetric tensor.

Now let us determine 
the lowest term of $W_\Delta$
in order to make the singular surface consistent with the discriminant 
of the M theory curve. 
If $W_\Delta = W_2 \equiv 
g_4 
f_2(z_3,z_2) \Lambda^6
$, then
\begin{eqnarray}
\langle u_2 \rangle
\!\!\!& =\!\!\!& 
z_3{}^2-z_2 \pm 2 \Lambda^3 G^{-1/2}
+ \frac{1}{g_4} \frac{\partial W_2}{\partial z_2}
\nonumber \\
\langle u_3 \rangle
\!\!\!& =\!\!\!& 
2 z_3{}^3 + 2 z_3 z_2 \pm 4 z_3 \Lambda^3 G^{-1/2}
+ \frac{1}{g_4} \frac{\partial W_2}{\partial z_3}
\nonumber \\
\langle u_4 \rangle
\!\!\!& =\!\!\!& 
-\frac{3}{2} z_3{}^4 -2 z_3{}^2 z_2 +\frac{1}{2} z_2{}^2
\pm 2 z_2 \Lambda^3 G^{-1/2}
- \frac{z_2}{g_4} \frac{\partial W_2}{\partial z_2}
- \frac{z_3}{g_4} \frac{\partial W_2}{\partial z_3}
+ \frac{W_2}{g_4}.
\label{2singular}
\end{eqnarray}
{}From this, we find 
\begin{eqnarray}
\lefteqn{\Delta(u_k = \langle u_k \rangle)} \nonumber \\
&\!\!\!=&\!\!\!
\mp 2048 \left( {z_3}^2+z_2 \right)^{3/2}
\left( 5 {z_3}^2+z_2 \right)^6
\left( f_2(z_3,z_2) 
\left( {z_3}^2+z_2 \right)
-2
\right) \Lambda^9
+ {\cal O}(\Lambda^{12}).
\end{eqnarray}
Thus, in order for $\Delta$ to vanish up to $\Lambda^9$, 
$W_2$ must take the form 
\begin{eqnarray}
W_2 &\!\!\! = &\!\!\! 
2 g_4
\left( {z_3}^2 + z_2 \right)^{-1} \Lambda^6
\nonumber \\
&\!\!\! = &\!\!\! 2 g_4 G^{-1} \Lambda^6.
\label{W2}
\end{eqnarray}
This $W_2$ (\ref{W2}) satisfies 
the conditions (\ref{WDeltacondition}) and 
carries the charge $(2,2)$.

We can determine the next term of $W_\Delta$
in a similar way and find
\begin{eqnarray}
W_d +W_\Delta
= 
\pm 4 g_4 \Lambda^3 G^{1/2}
+ 2 g_4 \Lambda^{6} G^{-1}
\mp 2 g_4 \Lambda^{9} G^{-5/2}
+{\cal O}(\Lambda^{12}).
\label{W3}
\end{eqnarray}

Inspired by the result (\ref{W3}),
we restrict the form of $W_{\Delta}$ in the following way
\begin{eqnarray}
W_d + W_\Delta = g_4 \sum_{k=1}^{\infty} h_k G^2
\left( \Lambda^3 G^{-3/2} \right)^k,
\qquad{\rm where}\qquad h_k \in {\bf C}.
\label{superpo}
\end{eqnarray}
By requiring for $\Delta$ to vanish, 
we work out $h_k$ up to $h_8$
\begin{eqnarray}
W_d +W_\Delta
&\!\!\! = &\!\!\!
\pm 4 g_4 \Lambda^3 G^{1/2}
+ 2 g_4 \Lambda^{6} G^{-1}
\mp 2 g_4 \Lambda^{9} G^{-5/2}
+ 4 g_4 \Lambda^{12} G^{-4}
\mp \frac{21}{2} g_4 \Lambda^{15} G^{-11/2}
\nonumber \\ &\!\!\! &\!\!\! {}
+ 32 g_4 \Lambda^{18} G^{-7}
\mp \frac{429}{4} g_4 \Lambda^{21} G^{-17/2}
+ 384 g_4 \Lambda^{24} G^{-10}
+{\cal O}(\Lambda^{27}),
\end{eqnarray}
and we find that the discriminant vanishes up to the 
order $\Lambda^{27}$: 
\begin{eqnarray}
\Delta(u_k = \langle u_k \rangle)= {\cal O}(\Lambda^{30}).
\end{eqnarray}

Although we have only determined $W_\Delta$ up to 
this order due to the increasing complexity of computation, 
we believe that the higher powers of $\Lambda$ can be worked out 
with more efforts and that the form (\ref{superpo}) will come out.

%
\vspace{7mm}
\pagebreak[3]
\addtocounter{section}{1}
\setcounter{equation}{0}
\setcounter{subsection}{0}
\addtocounter{footnote}{0}
\begin{center}
{\large {\bf \thesection. Discussion}}
\end{center}
\nopagebreak
\medskip
\nopagebreak
\hspace{3mm}

In this paper, we studied the singular surface of the moduli space of 
the $N=2$ SUSY $SU(N_c)$ gauge theory with an antisymmetric tensor 
matter from two points of view. 
One is to use a configuration of a single M theory five-brane and 
the other based on the ``integrating in'' method. 
It was discussed that the consistency between the two results requires 
$W_{\Delta} \ne 0$, and the explicit form of it was worked out
for $N_c=4$. 
It is interesting that 
$W_\Delta$ consists of terms with integer powers of $\pm \Lambda^3$ 
corresponding to two vacua of $SU(2)$ gaugino condensation. 
Using the dynamical scale $\Lambda_{SU(2)}$ of the unbroken 
$SU(2)$ SUSY Yang-Mills theory instead of $\Lambda$, the nonperturbative 
superpotential (\ref{superpo}) is rewritten as 
\begin{equation}
W_{d}+W_{\Delta}=
g_4 \sum_{k=1} h_k G^2 
\left( \pm {\Lambda_{SU(2)}^3 \over 2 g_4 G^2} \right)^k,
\end{equation}
where the scale matching condition 
$\Lambda_{SU(2)}^3=2 g_4 G^{1/2} \Lambda^3$
for $N_c=4$
is used. 
This fact makes us suspect that the physical origin
of $W_{\Delta}$ might be understood as the gaugino 
condensation of $SU(2)$ SUSY Yang-Mills theory. 

In order to break $N=2$ SUSY to $N=1$ SUSY,
we added a perturbation (\ref{3.1}) to the tree-level
$N=2$ superpotential $W_{\rm tree}$
in the ``integrating in'' methods.
Instead of this perturbation, we can consider another perturbation
\begin{eqnarray}
\Delta W 
= \frac{g_2}{2} {\rm Tr}(\Phi^2)
+ \frac{g_3}{3} {\rm Tr}(\Phi^3)
+ g_4 
\left(
\frac{1}{4} {\rm Tr}(\Phi^4) - \alpha
\left( \frac{1}{2} {\rm Tr}(\Phi^2) \right)^2
\right)
\qquad \alpha \in {\bf C}
\label{deltawwithalpha}
\end{eqnarray}
to the tree-level
$N=2$ superpotential $W_{\rm tree}$.
It turns out that for $\alpha \neq 1/2$ 
there exist classical vacua where the gauge group is broken to 
$SU(2) \times U(1)^2$ \cite{TY2}.
One can then calculate $W_\Delta$ for generic $\alpha$ by requiring 
that $\Delta=0$ in the $u_k = \langle u_k \rangle$ surface. 
One of the most interesting results is that
$W_\Delta$ vanishes 
for $\alpha=1/4$ \cite{private}
while $W_\Delta \neq 0$ for any other value of $\alpha$.
This is presumably related to the fact that the antisymmetric 
representation of $SU(4)$ is 
equivalent to the defining representation of $SO(6)$, for which 
the assumption $W_{\Delta}=0$ is found to be valid. 

It is interesting to note that the ``integrating in'' methods have been 
applied under the assumption $W_{\Delta} = 0$ \cite{EFGIR}-\cite{TY2}, 
which provides consistent results in the Seiberg-Witten curves that are 
hyper-elliptic. 
It has been known that the non-hyperelliptic Seiberg-Witten curves for 
the exceptional group cases are derived using the assumption $W_\Delta=0$ 
\cite{LPG}\cite{TY2}. 
Contrary to these results, 
we have found that the assumption $W_{\Delta} = 0$ is inconsistent 
in the case of $N=2$ SUSY $SU(N_c)$ gauge theory 
with an antisymmetric tensor matter, 
whose Seiberg-Witten curve is not of  hyper-elliptic type \cite{LL1}. 
As another example of nontrivial $W_{\Delta}$, 
we have studied also $N=2$ $SU(4)$ theory with a symmetric tensor whose
Seiberg-Witten curve is proposed in ref.~\cite{LL1}. 
In this case, we find that there are no complex number $\alpha$
in eq.(\ref{deltawwithalpha}) to make $W_\Delta$ vanish.
On the other hand, we also find that
for fundamental matters, 
$W_\Delta$ vanishes for any value of $\alpha$
($\alpha \neq 1/2$).

Higher values of $N_c$ may be dealt with by a similar method with
more computational efforts.
We expect that $W_\Delta \neq 0$ for $SU(N_c)$ with antisymmetric 
or symmetric representation for higher values of $N_c$ also.

\vspace{10mm}

We would like to thank T. Kitao 
for useful comments and a collaboration in the early unsuccessful 
attempts to obtain the curve. 
We
thank S. Terashima and S. K. Yang for 
illuminating discussion and comments, especially on the consistency 
of the non-hyperelliptic curve and $W_{\Delta}=0$ 
in various examples. 
The work of T.S. is supported by JSPS Research Fellowship for 
Young Scientists. 
This work is supported in part by Grant-in-Aid for Scientific 
Research from the Ministry of Education, Science and Culture for 
the Priority Area 291. 

\end{document}